# USING LHC AS INJECTOR AND POSSIBLE USES OF HERA MAGNETS/COILS

K. H. Meß, CERN, Geneva, Switzerland

*Abstract*

This workshop discusses the various aspects of a high energy version of the LHC in the LHC tunnel, the basic assumption being that the LHC will be decommissioned. The possibilities to recycle LHC and the already stopped HERA are discussed in this paper.

## INTRODUCTION

It might seem too early to discuss the fate of the LHC magnets before they have reached their design performance and well before the LHC has produced sufficient luminosity to support or change our present concept of high energy physics. However, ideas, like the HE-LHC, need a long time to be accepted, planned, and eventually transformed into reality. Trying to contain the costs by studying the possibilities of recycling high investments of the past is an integral part of this process.

By the time of the HE-LHC the LHC will be decommissioned and the superconducting magnets of HERA [1] in Hamburg might also still be available and useful, if the DESY management decides so.

Table 1 shows a summary of the parameters, which are most important for a recycle. The LHC has, as of today, not reached its design performance, while HERA has been operated in the last years about 12% above design (values in brackets). In both cases the magnets are optimised for the specific purpose. The magnets are bent to maximise the aperture while minimising the coil diameter (as well as cost and stored energy). Evidently, the curvature is adapted to the respective bending radius ρ, given by the magnetic field B and beam energy E:

$$\left| B\,[T] \right| = \frac{\beta \cdot E\,[GeV]}{0.2998 \cdot \rho\,[m]}$$

Table 1: Some LHC and HERA parameters

| Machine | LHC [2] | HERA [1] |
|---|---|---|
| Circumference | 26.7 km | 6.4 km |
| # of main bends | 1232 | 422 |
| Magnet length | 14.3 m | 8.9 m |
| Injection Field | 0.535 T | 0.227 T |
| Flat Top Field | 8.33 T | 4.649 (5.216) T |
| Current (inject.) | 763 A | 245 A |
| Current (top) | 11850 A | 5027 (5640) A |
| Inj.Energy | 450 GeV | 40 GeV |
| Top Energy | 7 TeV | 820 (920) GeV |
| Bending radius | 2804 m | 588 m |
| Inner coil Ø | 56 mm | 75 mm |
| Cold tube Ø | 50 mm | 55.3 mm |
| Sagitta | 9.14 mm | 14.4 mm |
| Nom. dI/dt | 10 A/s | 10 A/s |
| Tunnel Ø | 3.76 m | 5.20 m |

## LHC MAGNETS

The LHC magnets are designed for the LEP tunnel. Hence, for the useful range of the magnetic field, particle energy, and aperture, the magnets fit only into a tunnel of about 27 km circumference, i.e. the LEP tunnel.

### *Use of the LHC as injector*

The injection energy of the HE-LHC is planned to be around 1.3 TeV. A somewhat higher energy would of course be beneficial, both in terms of persistent current effects and total filling time. The "old" LHC could evidently accelerate from 450 GeV to anything below 7 TeV and keep the two beams ready for injection, provided that the beams do not interfere with the HE-LHC, while it is still running at a much higher energy. The LHC could prepare the beams "in the shadow" and shorten the overall filling time of the HE-LHC, despite its low acceleration rate. This scenario does not require a new SPS and new injection lines operating above 1 TeV.

### *Space requirements for the LHC and HE-LHC*

However, this forces the co-existence of the "old" and the "new" accelerator in a fair fraction of the tunnel and it needs new beam lines to bypass the experiments. The bypasses have to go through the galleries. To keep the two machines at the same length (which is essential for the use of the LHC as injector) the LHC has to be shifted towards the transport space everywhere else. Alternatively, the HE-LHC has to be shifted further to the outside, referred to the present layout. Neither of these options seems easily possible.

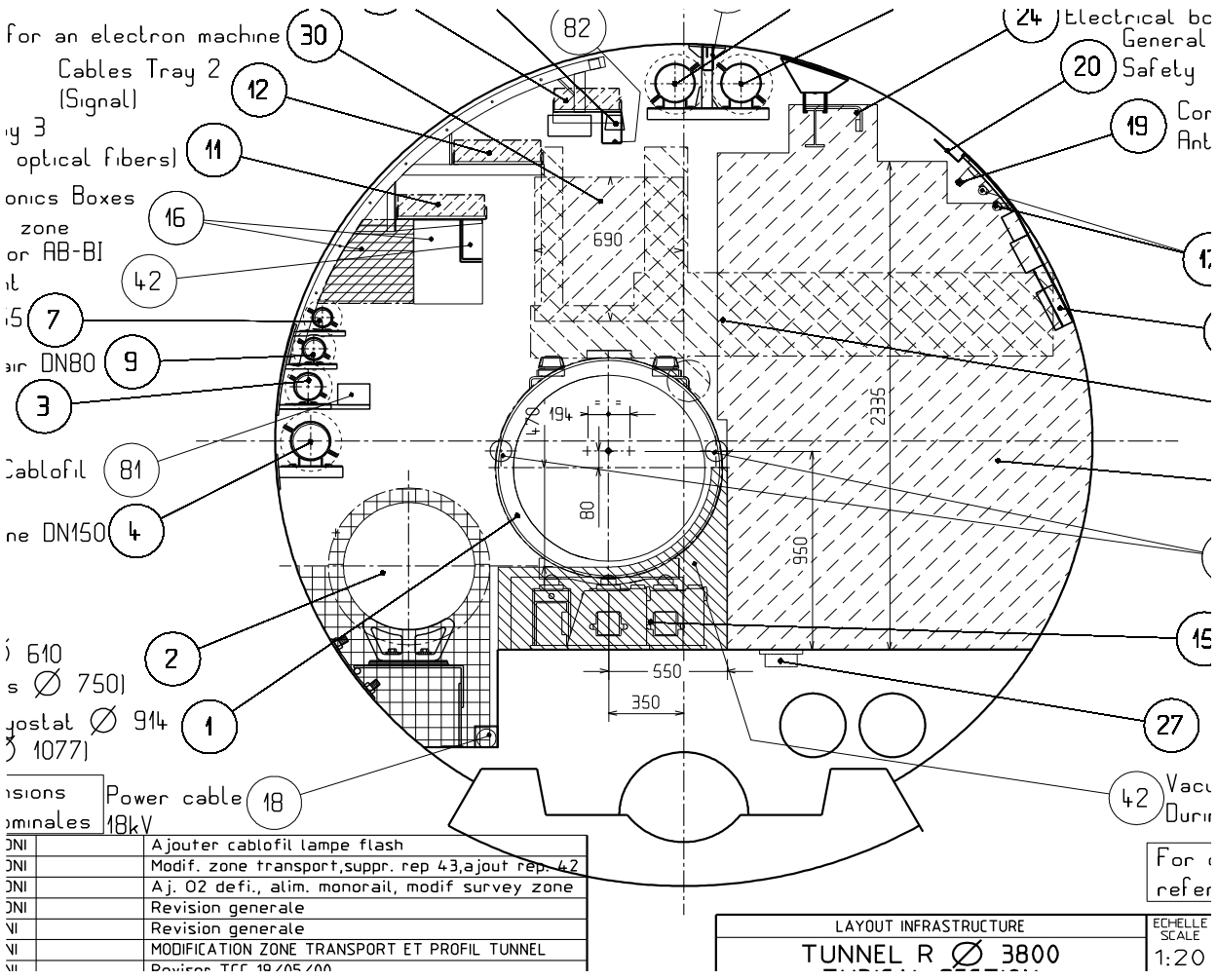

Figure 1. Sketch of the LHC tunnel

Fig 1 shows a sketch of the LEP tunnel with a LHC dipole. It presents the most benign case, without further obstructions in the way. It seems just possible to fit a much smaller electron accelerator of 50...60 GeV (LHeC) [3], [4]. The HE-LHC magnet is, however 230 mm wider than the LHC [5]. If the top energy of the LHC is decreased to, say, 2 TeV a large amount of flux steel can be taken away, thus saving space. The "slim" LHC and the HE-LHC could share the same cryostat. Such a combined machine might fit, but this proposal "works" only for the "easy" part of the machine.

Figure 2 and figure 3 show a more difficult case, where the trick, mentioned above, does not work: the area of the dump-ejection kickers and lines.

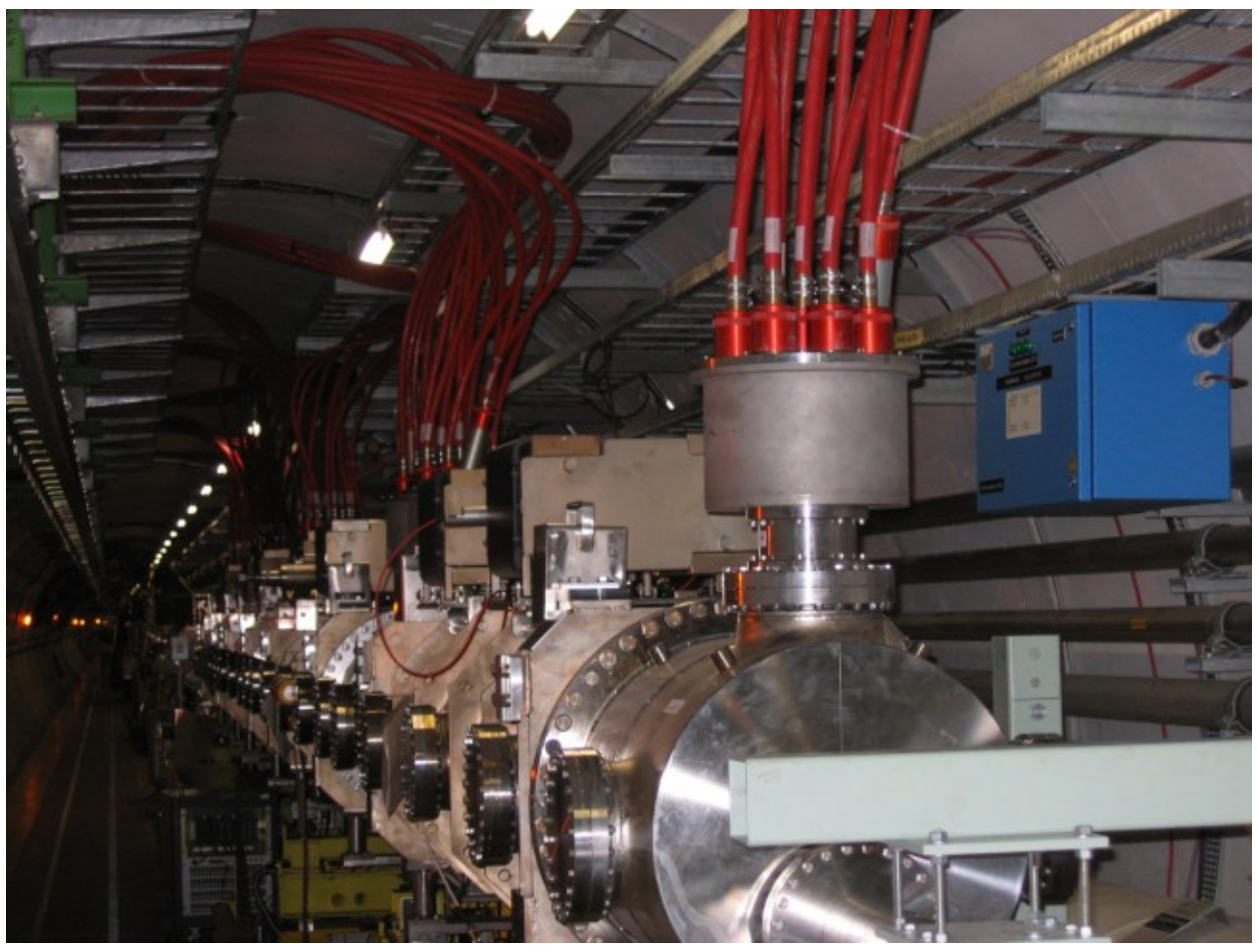

Figure 2. LHC dump kickers

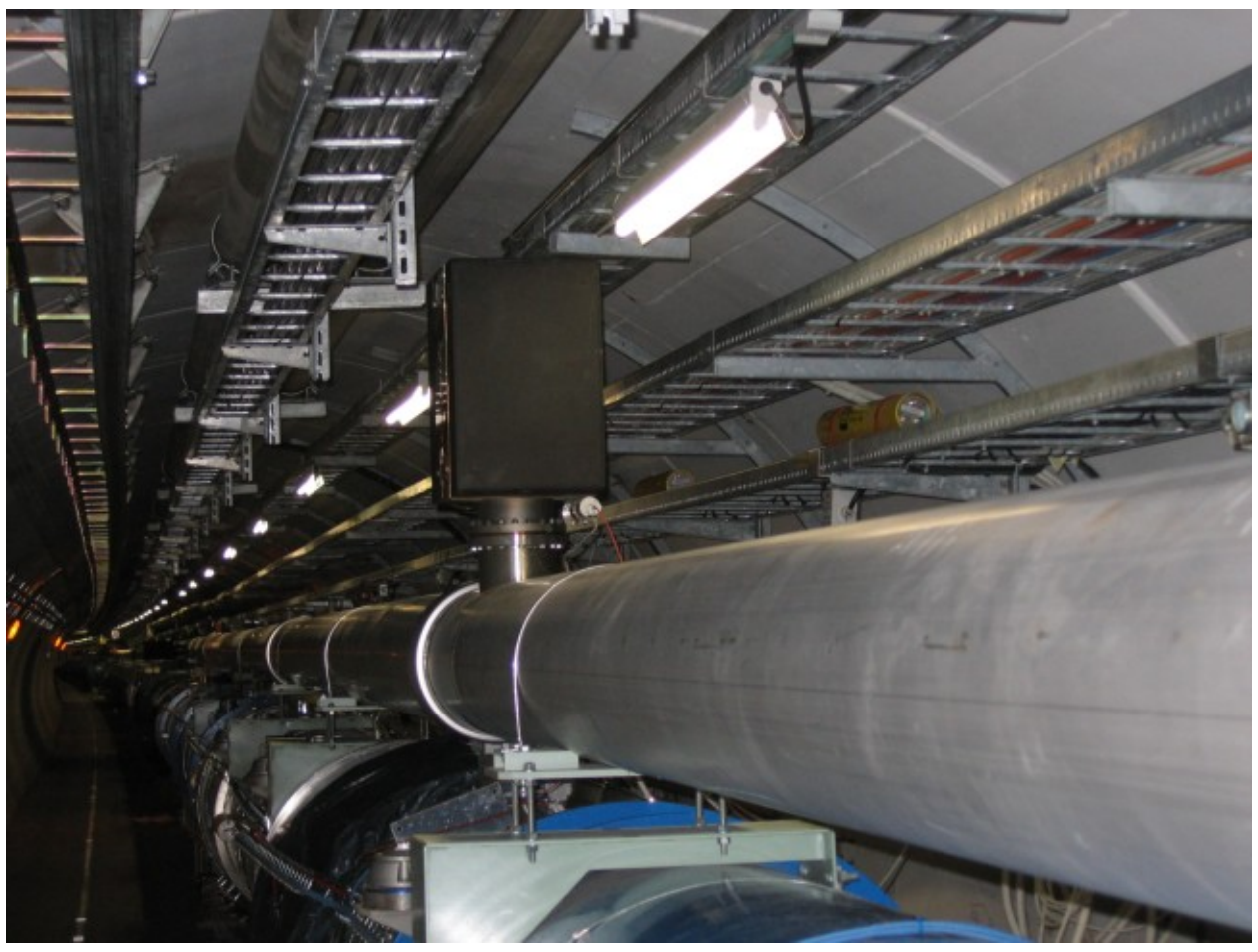

Figure 3. LHC dump line above the LHC

There is clearly no space to place a second set of kickers or dump lines as well as a wider HE-LHC, even under the assumption that kickers for 16 TeV can be produced fitting into the straight section longitudinally.

The RF section for the LHC would have to be moved to one of the new bypass sections, like it could be done for the LHeC. The cryogenics [6] for the HE-LHC, however, will be of the same size as the existing for the LHC. There is no space for it. The HE-LHC will need its own energy extraction system, which will consist of at least twice the number of switches and resistors as pointed out before [7]. There is no space for it. The HE-LHC will need its own set of collimators. The design is unknown [8]. However, the collimation system will not be smaller than the existing for the LHC. There is no space for it.

### *Conclusion*

The HE-LHC and the LHC cannot fit into the LEP tunnel. The LHC magnets cannot be re-used in the context of the HE-LHC.

## HERA MAGNETS

The case of the HERA magnets has been treated before [9]. This report summarises that work. Figure 4 shows a view of the HERA tunnel. The two accelerators are installed above of each other with the proton machine on the top. The dipoles [10] were produced partly in Italy (see figure) and partly in Germany. The quadrupoles were produced in France and mounted into their cryostat in Germany.

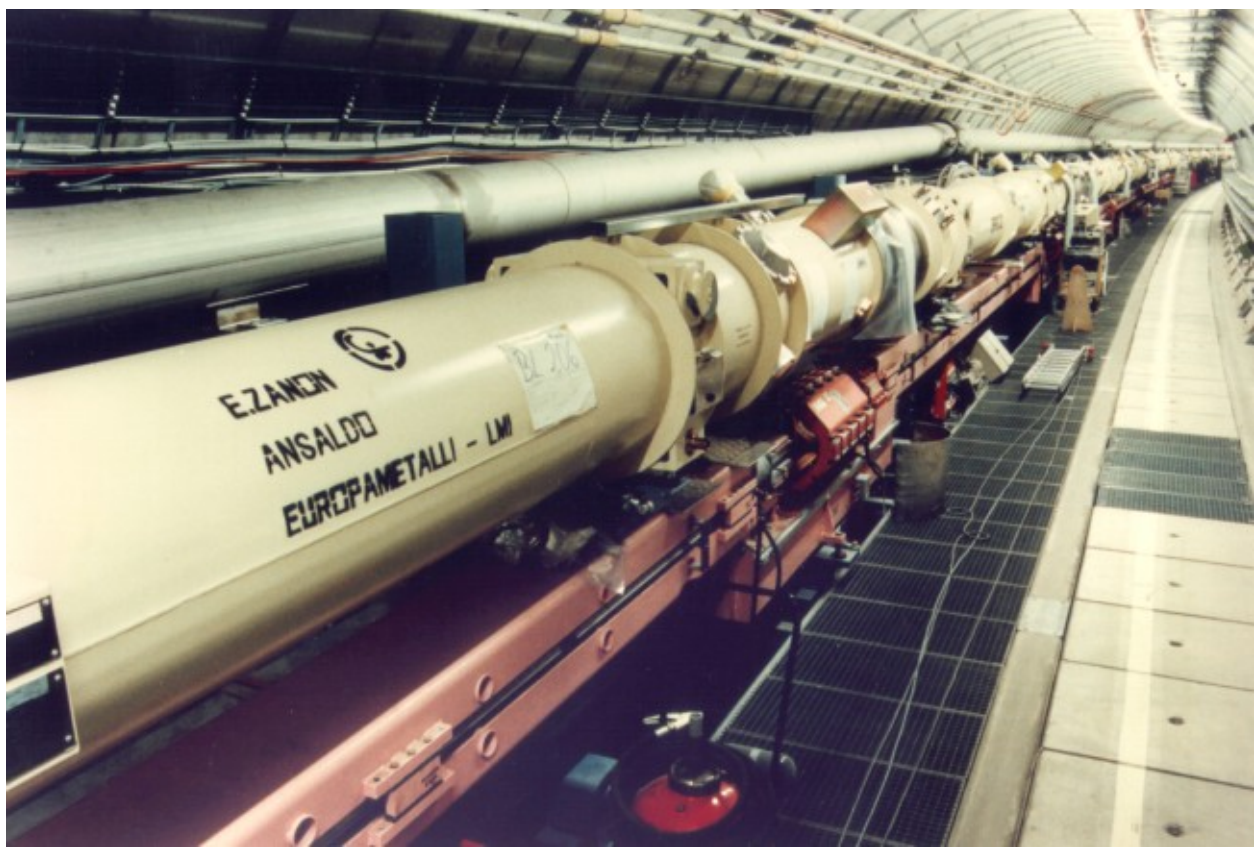

Figure 4. View of the HERA tunnel with the superconducting proton machine on top of the electron machine.

### *Use of the HERA magnets*

The HERA magnets are designed as storage ring magnets. Hence, the acceleration rate is low (~1.6 GeV/s). The use as pre-accelerator in the SPS tunnel is not attractive, although the radius of curvature could be adapted to. The SPS tunnel is wide enough and additional aperture could be created by replacing the beam pipe. The present beam pipe diameter is determined by the corrector windings on the beam pipe [11].

The slow acceleration rate is of no concern, if the magnets are used for the injection lines TI2 and TI8. The question is: do the magnets fit there and can the cryogenic requirements be met?

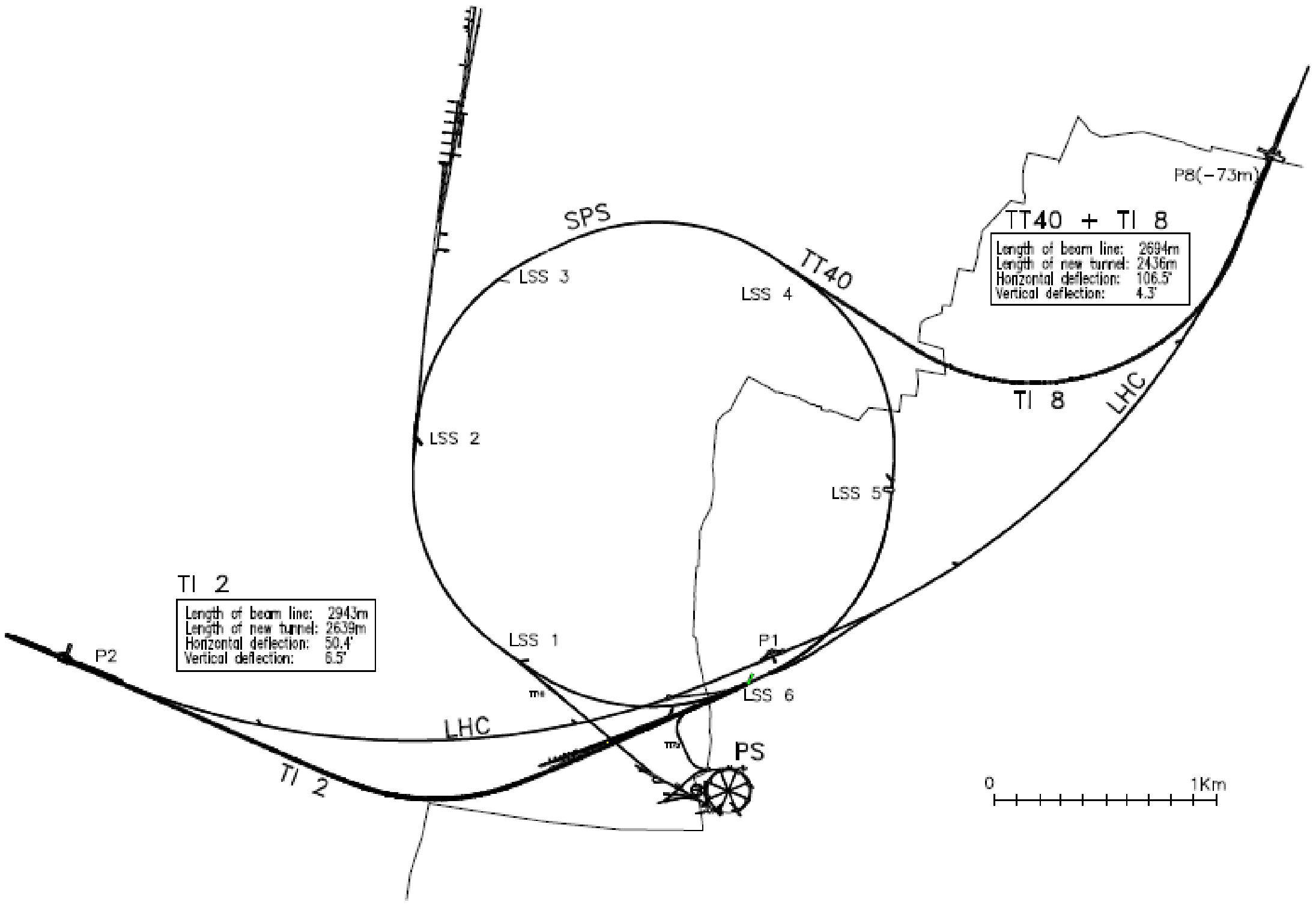

Figure 5. Layout of the injection lines TI2 and TI8

## *Injection line layout*

The layouts of the TI8 and TI2 transfer lines are described in the LHC Project Note 128 [12] and in the LHC Design Report, Vol III [2]. TI2 has a length of about 2.9 km. It consists of one 48° horizontal bend and three vertical bends of 61, 42 and 9 mrad to avoid the underground valley below St. Genis. The steepest slope is 2.6%. TI8 is somewhat shorter (2.7 km) but steeper (3.77%). It consists of a horizontal bend of 103° in the descending part, preceded and followed by vertical bends of 38 and 35 mrad respectively.

The lines use a FODO structure with a half-cell length of 30.3 m and 4 dipoles per half-cell for the horizontal bending part. The vertical bends are made of a different type of bending magnets. The main features of the injection lines are shown in Figures 5 to 7, taken from the LHC project note 128. Note that the proton beam is bent counter-clockwise in TI8 and clockwise in TI2. Note further that in both cases the magnets are placed at the inner radius of the injection tunnel.

## *The HERA magnets*

A HERA half-cell consists of one dipole on either side of the dipole-corrector and quadrupole assembly. The FODO cell has a length of 47.012 m. The dipoles contain beam-pipe corrector windings, as mentioned above. A dipole corrector and a beam position monitor are also integrated in the cryostat of the quadrupole. A few shorter quadrupoles and vertical dipoles exist to adjust the optics and to deflect the proton beam vertically. The key parameters of the various magnets can be found in Ref. [9].

In HERA the superconducting main magnets are connected in series. The current flows clockwise through the dipoles and counter-clockwise back through the quadrupoles. Hence the optical lattice is fixed. Adjustments to the tune are made by varying the relatively strong quadrupole correctors, wound around the beam pipe inside the dipoles. All dipoles are curved to follow the local bending radius of the beam of $r = 588$ m. The proton beam travels counter-clockwise in HERA. The magnets are placed on the outer side of the tunnel with the quench relief valves also pointing to the outside.

A HERA dipole deflects a 820 GeV beam by 2.9599 mrad at the nominal excitation with 5027 A (4.649 T).

The beam pipe is bent correspondingly. Note however that HERA has been operating for a number of years at 920 GeV with a field of 5.216 T. This was made possible by lowering the temperature of the coils.

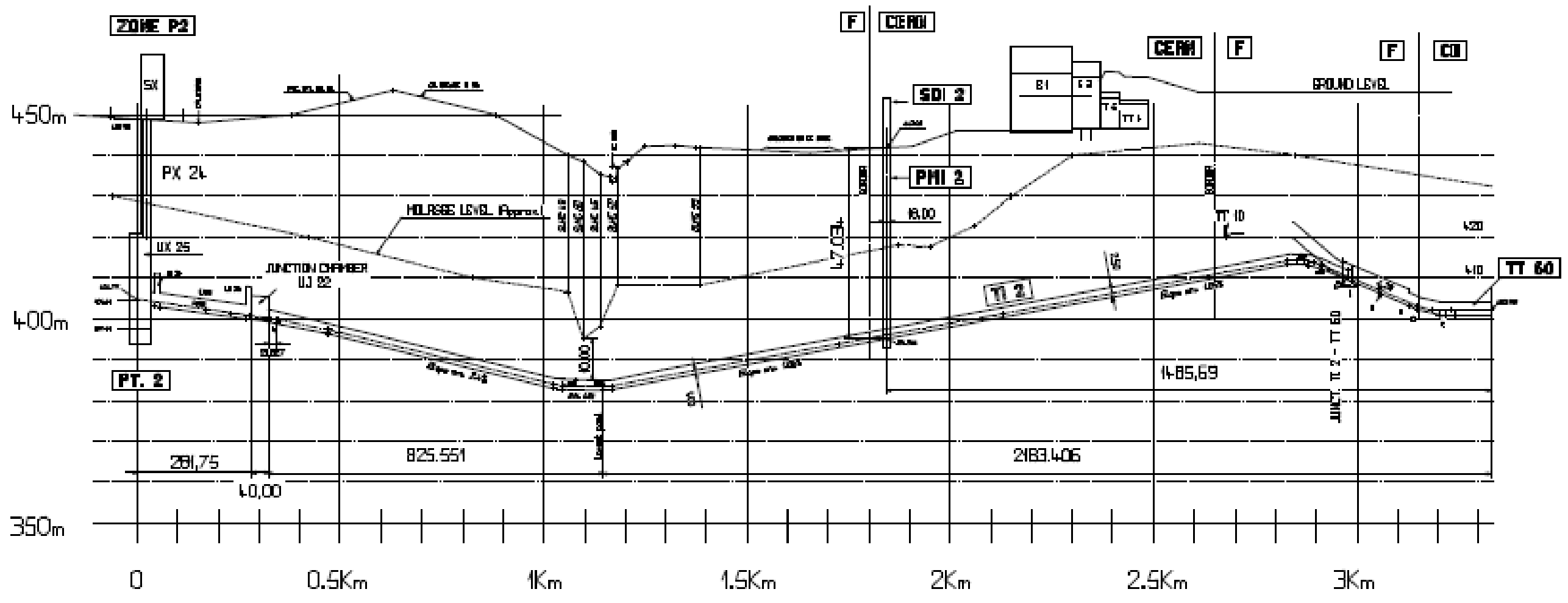

Figure 6. The vertical deflections in TI2

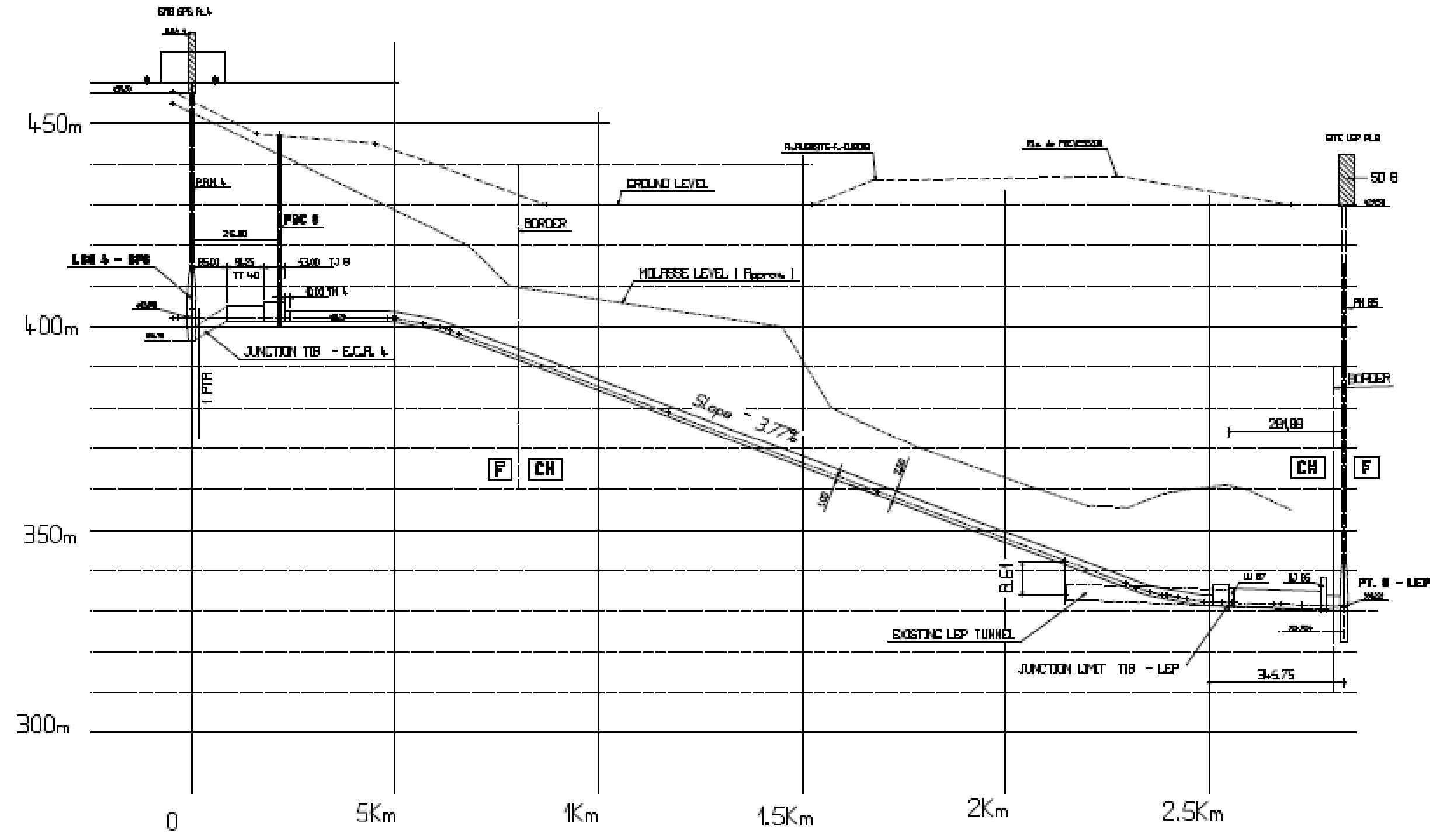

Figure 7. The vertical deflections in TI8

## *Space requirements and necessary changes*

The HERA tunnel is much wider than TI2 and TI8. It will be difficult to accommodate HERA magnets in the LHC injection channels. The situation is particularly difficult in TI2, which is also used to transport LHC dipoles into the LHC tunnel.

Fig. 8 shows on the left this situation as it is now and a possible solution with the HERA magnets at the right.

Note that the beam is presently on the inner curvature of the tunnel in both cases. The HERA dipoles have their quench relief valves and the quench exhaust pipe at the outer curvature (i.e. in the transport space in this case), which would clearly obstruct the transport zone. Either the cryostats have to be modified or the beam line has to be moved to the outer curvature. The latter is not easy, because the position of PMI2 was chosen to lower LHC magnets into the space at the outer curvature of TI2. One could presumably install a transfer table at the lower end of the shaft, such that the TI2 magnets are in fact installed underneath the shaft at the outside curvature. Components for the LHC could then be lowered to the transfer table and moved sideways and lowered into TI2, to pass on the inner curvature. In this way one could avoid dismounting the vacuum pipe, which presently blocks the transport path. Fig. 8 does not show any cryogenic line. The number of cable trays, however, cannot be reduced drastically (it is likely to increase, because the quadrupoles need cables).

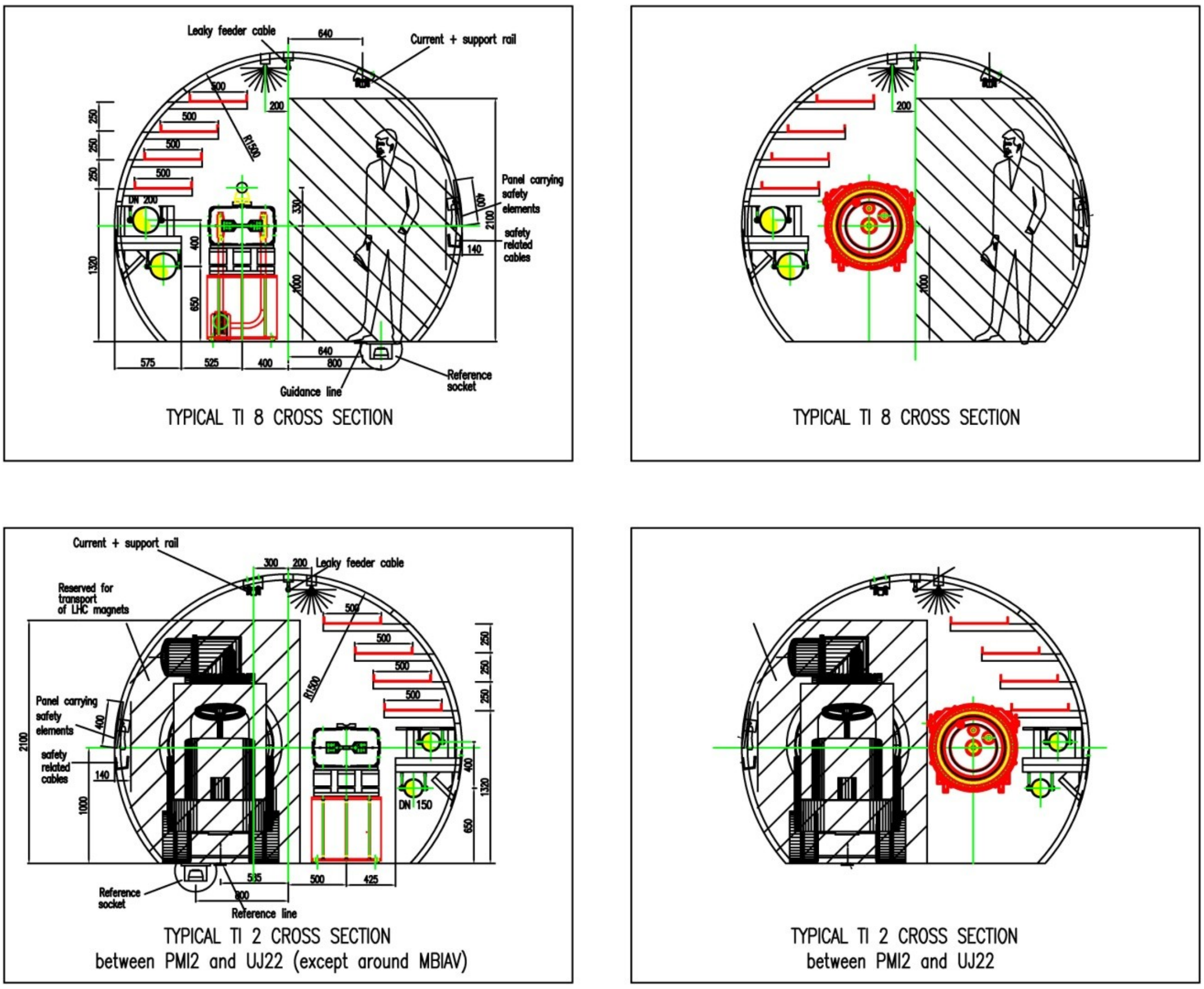

Figure 8. The TI2 and TI8 cross-sections in the present state (left) and with HERA dipoles (right)

It seems extremely difficult to fit the HERA magnets into TI2, as is illustrated by Fig. 9, now for the magnets on the outer curvature. The 500 mm quench relief pipe does not fit. The quench relief valves would need to be seriously reworked and still the cables would not find space. Even worse, as can be seen in Fig. 6, the deepest point of TI2 is underneath the creek Lion. According to the studies on cryogenics for the injection lines (see below), the single-phase helium will have to enter the string of magnets at the lowest point. However, the study does not take the actual configuration into account and will have to be repeated.

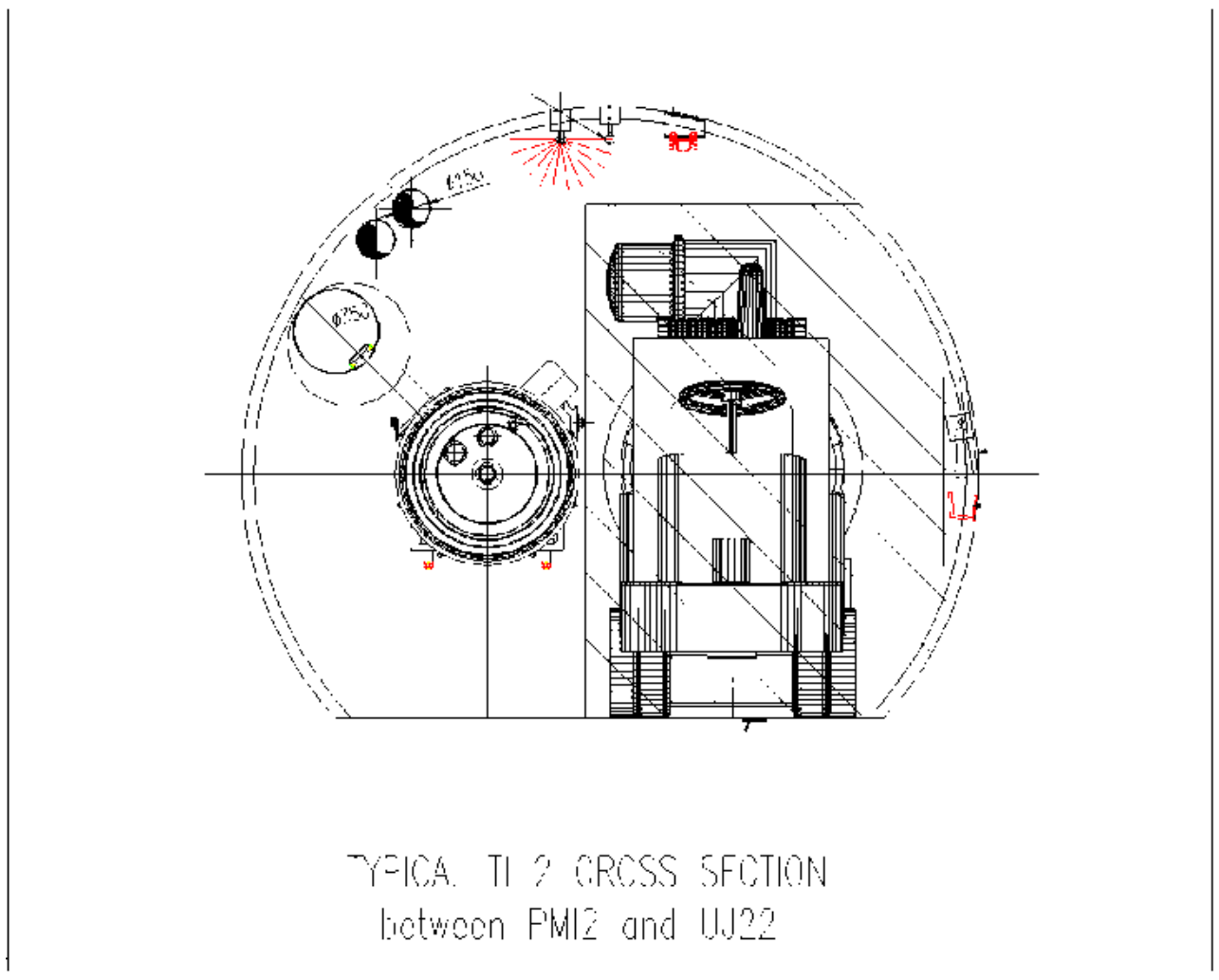

Figure 9. HERA dipoles in TI2 on the outer curvature in beam direction (two versions of the quench relief pipe)

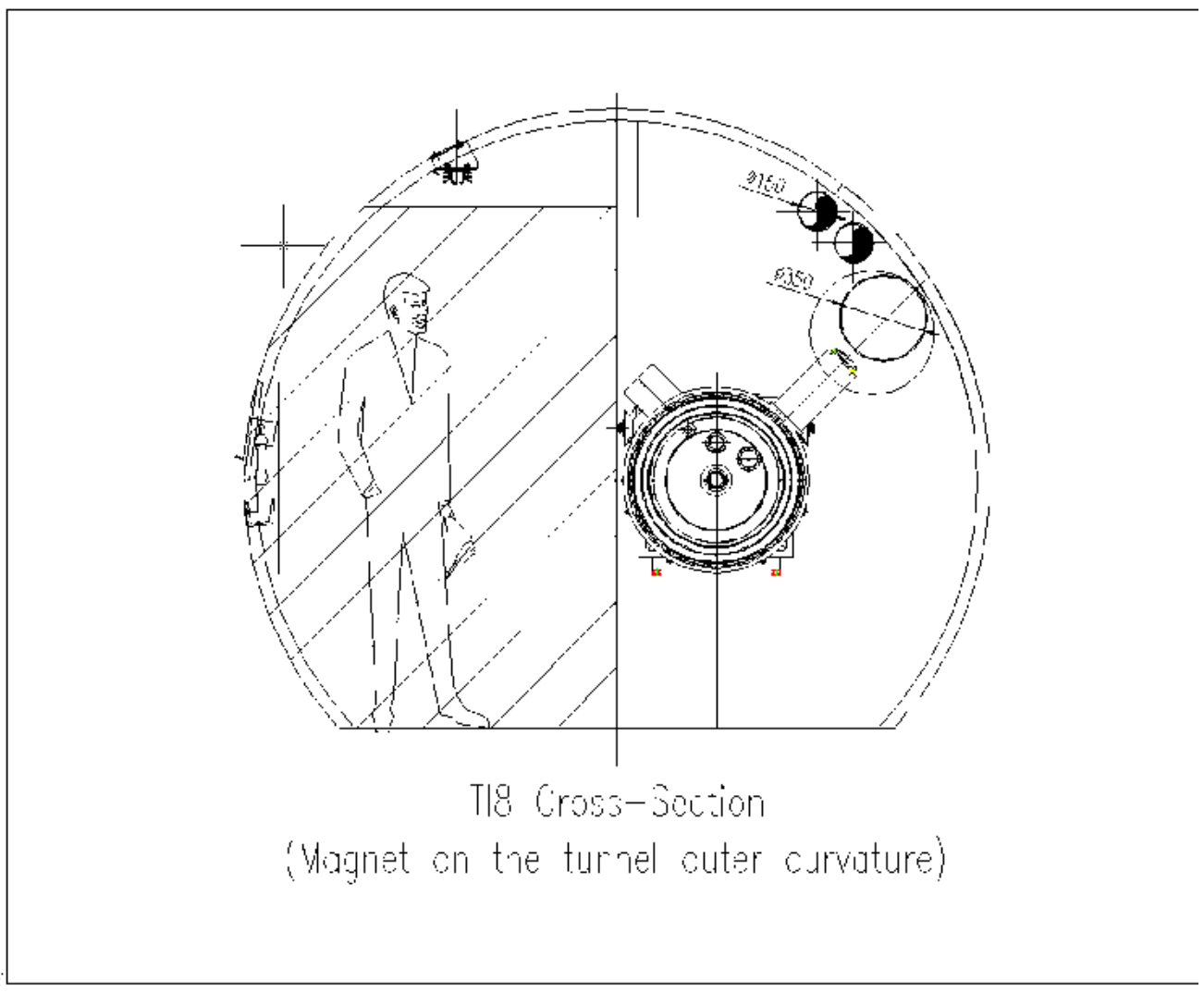

Figure 10. HERA dipoles in TI8 on the outer curvature in beam direction.

In TI2 the bending is clockwise. The HERA magnets would need to run with inverted polarity, which requires a change of the polarity of the protection diode. This can be achieved by removing the diode stacks, opening them, inverting the polarity, testing them under cryogenic conditions and reassembling. This is a tedious, but possible, operation. Alternatively, adapter pieces could be envisaged, which change the polarity inside the cryostat. This seems possible, because both magnet ends (end covers) will have to be opened in order to fulfil the conditions set by the cryogenics (see below).

Still, the cable ladders do not find space and water cooling will have to move also. While the cables are still needed, the water cooling could maybe be reduced. Presumably the beam line could be lowered somewhat, which seems possible at this stage. It is unclear, if and where the cryogenic re-coolers could be placed, unless they can be part of the connection cryostats or the magnets themselves.

The situation is slightly better in TI8. The bend is counter-clockwise, as in HERA. Hence the magnets can run with the original polarity. The magnets are however also here on the wrong side of the tunnel. Again they would have to be moved to the outer curvature to give space to the quench relief valves and quench pipe, which looks impossible, as can be seen in Fig. 10. Certainly the exhaust valves ("Kautzky valves") need to be reworked and the beam line has to be lowered.

### *Limits*

It might be possible to rearrange the optics [11] to make optimal use of the properties of the HERA magnets and achieve higher energies. However the HERA magnets can also be mapped onto the existing structure. Because of the higher bending power of the HERA dipoles, compared to the normal conducting magnets, the present cell length of 60.6 m is sufficient to reach 900 GeV. The space between the dipoles will be filled with connection cryostats, containing the quadrupoles, the current leads and cryogenic feed-boxes. The limitation to 900 GeV is given by two constraints: the optics chosen as baseline and the bending radius of the magnets. Both constraints are somewhat flexible. The density of dipoles could be increased and hence the total bending power. However, in this case, the aperture would be reduced, due to the poorly matching sagitta of the beam pipe. This seems acceptable comparing the HERA aperture with the present beam line aperture. In conclusion, a 1 TeV beam line of sufficient aperture could presumably be made with a new optics design.

The proposed structure has, however, a very serious problem. In HERA the dipoles and quadrupoles are connected in series, containing only one bus-bar pair in the bypass. This is incompatible with the existing optics in TI2 and TI8 and the corrector quadrupoles inside the dipoles (2* 10.62 T integrated gradient) are insufficient to replace real lattice quadrupoles. They could, however be used for adjusting the optics.

Reference 9 lists all required dipoles and the aperture mismatches, all of which are very small. It also shows that the quadrupoles of HERA do not fit at all. New quadrupoles would be needed. MQTL [2] type magnets would be sufficient, limiting the required current to below 600 A. The current could be distributed using the blocked cooling channel or by adding a “line N” type pipe on the outside of the cold mass.

## Cryogenics

At present HERA is running under the following cryogenic conditions [13], [14]:

The magnets are cooled with supercritical helium with p> 2.5 bar, T=4.0 K (!). The supercritical helium is cooled by the counter flow of two-phase helium. The expansion is done at the lowest point of an octant, which is in the middle or at one end of an octant. There is a feed box containing the current leads and all the valves including a Joule Thomson valve. The two-phase flow is always directed uphill to avoid the capture of bubbles. To run at 920 GeV at HERA (as would also be required in the LHC injection lines) the temperature must be as low as T=4.0K. This is achieved by lowering the suction pressure of the screw compressors to 650 mbar. The pressure drop over the 620 m of one octant is about 100 mbar. The inclination of the HERA tunnel (max 10 mrad) is so small that the resulting pressure drop due to gravity can be neglected. The stationary mass flow is 35 g/s. A study has to be made of how to achieve similar conditions in the steep LHC transfer line tunnels.

In 1993 N. Delruelle et al. [15] studied a possible cryogenic system for the injection lines. At that time the ideas about the injection lines had not yet converged to the present design. Hence not all conclusions in this study can be applied to the present case. However, the slopes were planned to be even higher. The authors assume HERA or UNK like magnets of only 5.7 m length at 4.5 K. The preferred solution foresees single phase helium with re-coolers. The helium is fed in at the bottom of the arc (of which 3 were planned at that time) and proceeds through the magnets at a rate of 60 g/s. The liquid is re-cooled at the end of each cell by a heat exchanger in a bath of boiling helium. The gas is returned through the magnets using the holes in the iron laminations of the magnets, which in the case of the HERA magnets is either used for the heat exchanger or blocked (lower orifice). Thermally insulated pipes have to be inserted, to prevent heat propagation between the cells. Note that HERA quadrupoles do not have these heat exchanger holes, a further reason, why they cannot be used here. In addition a 500 mm quench relief line is needed.

Alternatively a two-phase cooling scheme with phase separators at the end of the cells has been considered. This scheme offers many advantages. However the authors request further tests before the solution can be seriously pursued, because “its feasibility is still doubtful”.

The study does not include the very special actual geometry of TI2 with its up-hill and down-hill slopes. The narrow tunnel will not be able to accommodate a refrigerator. A study has to be made on the basis of the actual geometries, whether and how stable conditions can be achieved at 4.0 K.

## Protection issues

The HERA dipoles come with a quench protection system [16], which is based on magnetic amplifiers. To operate this system requires special know-how, which is difficult to find even now. It has to be replaced in due time. Likewise, the capacitors in the heater power supplies will not operate any more after 40 years. In short, the electronics needs to be replaced.

The magnets are protected against the energy of the other magnets in the string by cold diodes. The diodes and the heat sinks are constructed to survive a decay time of 20 s from 6 kA. As the maximum voltage during the extraction has to be limited to below ±530 V, leading to an extraction resistance of less than 175 mΩ, the maximum inductance per protection block is limited to 3.3 H or 55 magnets of 60 mH each. This is close to the 56 or two times 57 magnets needed in the long arcs, but a bit too low. The resistors in HERA are simple bifilar stainless steel pipes, which could be reused adding some electrical protection. The switches are laterally of the size of the magnets and should fit. The same holds for the electronics.

A number of dipoles will need its own power converter. In these cases the diode might prevent a fast discharge.

The quench protection and energy extraction for the quadrupoles depend on the choice of the quadrupole system. Forty quadrupoles of the MQTL type connected in series have an inductance of 5 H. Bypass resistors, as implemented in the LHC for this magnet type, will be necessary. The resulting time lag is not important for the application as injection line magnets.

## Conclusions

The special geometry of the TI2 and TI8 transfer lines poses serious problems for upgrading them into the 1 TeV range. The HERA dipoles, with the required cryogenic pipe and cable trays, will not fit in, unless heavily reworked. A major rework of the magnets is also necessary to accommodate to the different cryogenic conditions (and the opposite field direction in TI2). The magnets will have to be taken out of their cryostat, the end-covers will have to be removed, the heat-exchanger pipes will have to be replaced and new connections will be necessary. Eventually only the collared coils with their flux iron can be reused. In any case, the end covers need to be closed again, after rerouting the pipe for the exhaust valve. Finally, new cryostats will have to be constructed.

The HERA quadrupoles can in all probability not be used, unless the optics is completely changed and the cryostats, the cold bus-bars and the internal helium pipes are redesigned. As a result around 180 new quadrupoles will have to be made.

The cryogenics has to be extraordinary slim in order to fit into the tunnel. The steep slope puts additional constraints. In particular TI2 with positive and negative slopes presents problems. This should be addressed in a separate study again.

## SUMMARY

In summary, the LHC and HE-LHC together do not fit in the LEP tunnel. The curvature of the LHC dipoles prevents any other use (except very special cases and in small quantities).

HERA dipoles could be used for a 900 GeV, probably 1TeV, beam line in TI2 and TI8. This requires however important changes of the cooling scheme and consequently of the cryostats. The quadrupoles cannot be used.

In view of these difficulties the use of combined function magnets [17] might have more advantages.

The cooling scheme has to be designed yet and may be very space consuming. This applies, of course, for any kind of superconducting injection lines.